\documentclass[twocolumn, prd, aps, nofootinbib, showpacs,showkeys, 
] {revtex4-1}

\pdfoutput=1
\usepackage{epsf,epsfig,subfigure,graphicx,amsmath,amssymb}
\usepackage{color}
\usepackage{dcolumn}
\usepackage{hyperref}

\def\lsim{\mathrel{\rlap{\lower4pt\hbox{\hskip1pt$\sim$}}
    \raise1pt\hbox{$<$}}}         
\def\gsim{\mathrel{\rlap{\lower4pt\hbox{\hskip1pt$\sim$}}
    \raise1pt\hbox{$>$}}}         

\newcommand{\red}{\textcolor{black}}

\begin{document}

\title{Superheavy dark matter and IceCube neutrino signals:\\bounds on decaying dark matter}

\author{Carsten Rott$^{1}$ \thanks{adbvdddd}, Kazunori Kohri$^{2,3}$, Seong Chan Park$^{4,5}$}
\email{rott@skku.edu, kohri@post.kek.jp, sc.park@yonsei.ac.kr}

\address{${}^1$Department of Physics, Sungkyunkwan University, Suwon 440-746, Korea\\
${}^2$Theory Center, Institute of Particle and Nuclear Studies,
KEK (High Energy Accelerator Research Organization), 1-1 Oho, Tsukuba 305-0801, Japan \\
${}^3$The Graduate University for Advanced Studies (Sokendai), 1-1 Oho, Tsukuba 305-0801, Japan\\
${}^4$ Department of Physics and IPAP, Yonsei University, Seoul 120-749, Korea\\
${}^5$ Korea Institute for Advanced Study, Seoul 130-722, Korea
}

\vspace{1.0cm}
\begin{abstract}
Superheavy dark matter may show its presence in high energy neutrino signals detected on earth. From the latest results of IceCube, we could set the strongest lower bound on the lifetime of dark matter beyond 100 TeV around $10^{28} {\rm sec}$. The excess around a PeV is noticed and may be interpreted as the first signal of DM even though further confirmation and dedicated searches are invited. 
\end{abstract}



\preprint{KIAS-P14052}

\maketitle


\section{Introduction}

One of the most pressing problems in nature is to understand the origin of dark matter (DM) and measure its properties~\cite{DM-Review}.  Several DM  candidates have been suggested but weakly interacting massive particles (WIMPs) have attracted the largest attention since WIMPs can naturally explain the observed density of DM thanks to the effective `WIMP miracle' following the Lee-Weinberg calculation in big bang theory~\cite{Lee:1977ua,Steigman:2012nb}. A relatively low mass window below $100$ TeV is open for WIMPs because of  the model independent theoretical upper bound coming from {\it perturbative unitarity}~\cite{Griest:1989wd} thus main experimental efforts for DM searches have been given to this low mass window.  No confirmed experimental evidence, however, has been found after many years of endeavour~\footnote{ For WIMP DM below TeV, the most stringent bounds on the spin independent DM-nucleon cross-section have been obtained $\sigma_{\chi N} \leq 2.0 (7.6) \times 10^{-45 (-46)} {\rm cm^2}$ by XENON100 (LUX) experiment at the WIMP mass $M_\chi =55 (33)$ GeV~\cite{Aprile:2012nq,Akerib:2013tjd}. It needs to be further improved to the level of $\sigma_{\chi N} \sim 10^{-49} {\rm cm^2}$ in the future to cover all the relevant parameters for WIMP in minimal supersymmetric standard model.
}. On the other hand,  the heavier mass regime beyond the unitarity bound has attracted less attention  even though superheavy DM, or WIMPZILLA, could be produced by non-thermal processes and explains the observed DM density in universe~\cite{WIMPZILLA}.  For strongly interacting superheavy DM, see~\cite{SIMPZILLA}.  DM could be also produced in high energy inelastic scattering processes~\cite{Harigaya}. The stability of superheavy DM was discussed earlier~\cite{Hamaguchi:1998wm, Park:2013bza}. 

Differently from DM at TeV scales~\cite{Ackermann:2012rg, Cirelli:2013hv}
, superheavy DM candidates are hard to test because
\begin{itemize}
\item the high mass surpasses the currently available collider energies so that DM production is kinematically forbidden,  
\item the longevity of DM implies that interaction strengths with the standard model particles are largely suppressed. Even worse, the number density in the galaxy is low ($\sim 1/M_\chi$) so that the direct detection rate by a detector on earth becomes doubly suppressed, where $M_\chi$ is the dark matter mass.
\end{itemize}

However, if DM decays,  the decay products carry distinctively high energies $E \sim M_\chi /N$ when the DM turns into a small number ($N$) of particles unless they are highly red shifted~\cite{Moroi}.  This opens a unique new window for superheavy DM.  In particular, neutrino among other potential decay products has advantage as a messenger particle as it can be directly observable on Earth preserving initial properties of DM. 

Indeed, IceCube recently reported their observation of high energy neutrinos in 30-2000~TeV~\cite{Aartsen:2013jdh, Aartsen:2013bka, Aartsen:2014gkd}. Very interestingly, the observed neutrinos are isotropic in arrival directions and show no particular pattern identified in arrival times, which suggest that the source is not local and violent but broadly distributed and stable~\cite{Aartsen:2014gkd}, which is consistent with the properties of DM \cite{yanagida, Serpico}:  stable (lifetime $\gg 4.3\times 10^{17}{\rm sec}$) and broadly distributed DM in the Galactic halo. 

The main goal of this letter is to examine the IceCube results and what they can tell us about superheavy DM with masses $M_\chi>100$ TeV. In Sec.\ref{sec:model} we show that decaying DM provides better fit to the IceCube data compared to annihilating DM then provide a benchmark model for concrete analysis. We set the most stringent new bound on the lifetime of DM above 100~TeV in Sec.\ref{sec:bound}. We note that some excess events in PeV-energy could be interpreted as an indication of decaying DM, which needs further studies. We conclude in Sec.\ref{sec:conclusion}.

\section{A benchmark model of Decaying DM}
\label{sec:model}

Two key observables that could distinguish decaying DM from self-annihilating are directional information and energy of neutrino signals. The neutrino flux is proportional to the density ($\rho$) or the square of the density ($\rho^2$) for decaying and annihilating DM, respectively.  As a consequence, the signals are more localized toward the Galactic center (GC) for annihilating DM but more isotropic for decaying DM. Assuming the NFW profile~\cite{Navarro:1995iw} we expect more than $50\%$ of events are within $65^\circ$ from GC for decaying DM but  within $25^\circ$ for annihilating DM.  Isotropy of the observed neutrinos would prefer the decaying DM interpretation. As the observed neutrino energies surpass 100~TeV, annihilating WIMP DM is excluded as a source because the energy domain lies beyond  the perturbative unitarity bound.

A model of decaying DM producing neutrinos by $\chi \rightarrow \nu h$ is suggested in accordance with the seesaw mechanism. We introduce a Majorana DM component in addition to the original seesaw Lagrangian and extend  the mass matrix by a small non-diagonal part, which eventually leads to a small coupling of DM to the neutrino and the Higgs boson.  
The model Lagrangian is given as 
\begin{eqnarray}
{\cal L} \ni -\lambda \overline{\nu_L}(h+v) n  -(\overline{n^c},\overline{\psi^c}) \begin{pmatrix}
M_n  &\sigma \\ 
\sigma & M_\psi 
\end{pmatrix} \binom{n}{\psi},
\end{eqnarray}
where $v\approx 246 {\rm GeV}$ is the vacuum expectation value of the Higgs field. Here we assume $M_n>M_n-M_\psi > M_\psi \gg \sigma$ and a negligibly small Yukawa coupling with $\psi$. Flavour indices are suppressed. The mass matrix is first diagonalized with the eigenmasses $M_\pm= \frac{1}{2}(M_n+M_\psi \pm \sqrt{(M_n-M_\psi)^2+4\sigma^2})$ and the mass eigenstates  
\begin{eqnarray}
\chi_+ &=& \cos\theta n+\sin\theta \psi, \\
\chi_-&=& -\sin\theta n +\cos\theta \psi. \nonumber
\end{eqnarray}
The mixing angle is $\theta= \tan^{-1} (\sigma/2\delta) \ll 1$ where $\delta=\frac{1}{2}\left(M_n-M_\psi \right) \gg \sigma$ as we assumed.  In terms of $\chi_+$ and $\chi_-$, the interaction Lagrangian is rewritten as
\begin{eqnarray}
{\cal L}_{\rm int}&=&-\lambda h \overline{\nu}_L n \nonumber \\
&\approx&  -\lambda h \overline{\nu_L} \left( \chi_+ + \frac{\sigma }{2\delta}  \chi_-\right).
\end{eqnarray}

For DM, $\chi_-$, the suppressed effective coupling constant $\lambda_{\rm eff}=\lambda \cdot\frac{\sigma}{2\delta}$ provides the largely suppressed decay width%
\begin{eqnarray}
\Gamma_{\chi_-\to \nu_L+h} &\approx & \frac{\lambda_{\rm eff}^2}{32\pi}M_-,
\end{eqnarray}
that provides the required lifetime to account the excess at PeV, $\tau \simeq 1.9 N_\nu \times 10^{28} {\rm s} $~\cite{yanagida},  with $\lambda_{\rm eff} \simeq 5.3\times 10^{-29}$. A large suppression factor is provided by a small mixing mass $\sigma\sim 10^{-5} {\rm eV}$  when the seesaw relation $m_\nu = \frac{(\lambda v)^2}{M_+}$ is assumed with $M_+\sim M_{\rm GUT} \sim 10^{14}{\rm GeV}$ and $m_\nu =0.1 {\rm eV}$. 
We identify $\chi_-$ as dark matter candidate and $M_-=M_\chi$.

\section{Bounds on decaying DM}
\label{sec:bound}

In this section we review existing bounds on heavy decaying DM and derive the most stringent limit for DM masses above 100~TeV based on recent public IceCube data.~\cite{Aartsen:2014gkd}

Neutrinos are often described as the least detectable channel, however for heavy decaying DM signals this picture changes and neutrinos turn into the most competitive detection channel. This can be understood by the fact that neutrino backgrounds are steeply falling as function of energy and that the neutrino cross-section increases with energy compensating for reduced number densities of DM particles. IceCube has produce a very stringent bound on DM decays in the Galactic halo using one year of data collected with the the partially instrumented detector. The limit on the lifetime for a heavy particle decaying into two neutrinos is strongest at the largest mass considered of 100~TeV and given with $10^{27}{\rm s}$~\cite{Abbasi:2011eq}. Searches with anti-protons and gamma-rays have also produced limits on decaying dark matter with masses up to 5~TeV and 10~TeV, respectively. Depending on the decay channel lifetime limits are typically between $10^{25}$~s to $10^{27}$~s.  
Cosmological constraints have been set~\cite{Gondolo:1991rn} and limits derived from neutrino data in scenarios with $\chi \to \nu\nu$ and $\chi \to ee\nu$ ~\cite{PalomaresRuiz:2007ry,Esmaili:2012us}.

In our analysis we consider a different decay process with one neutrino in the final state. In perspective to previous results we can improve upon them by taking a larger datasets using three years of the completed IceCube detector and by focusing on the most detectable signal orginating from the highest energy neutrino flux. We do not include the continuum component of the neutrino spectrum of $\chi \rightarrow h \nu$ as it would yield at most as many observed neutrino events as expected from the line. This can be understood from the fact that the neutrino cross section scales linear in $E_{\nu}$ and hence lower energy continuums neutrinos would come at the price of a lower interaction cross section. They further are spread over a large energy range requiring to introduce additional selection criteria to distinguish them better from backgrounds. By not introducing any angular selection criteria perform the analysis halo model independent.
For a recent review on these bounds we refer the reader to the CF2 Snowmass  working group summary~\cite{Buckley:2013bha}. Murase et al.~\cite{Murase:2012xs} compared sensitivities in gamma-rays and neutrinos for various dark matter annihilation and decay channels showing that neutrinos are most competitive for high DM masses.

IceCube recently reported the observation of a high-energy extra-terrestrial neutrino flux~\cite{Aartsen:2014gkd}. The data from this result can be used to also set a limit on the lifetime of heavy dark matter. We present a straight forward analysis that produces a conservative bound based on the data and invite the collaboration to perform a dedicated analysis to improve on our derived limit.

We assume that the dark matter distribution in our Galaxy follows an NFW profile and that decays from this Milky Way halo dominate compared to extra-galactic contributions, that are neglected. We note that for dark matter decays the choice of halo profile has a rather small impact on the expected flux. Previous works found that the extra-galactic contribution is about one order of magnitude smaller compared to the Galactic flux~\cite{Beacom:2006tt,Yuksel:2007ac,Yuksel:2007dr}.

\begin{figure}[t]
\begin{center}
\includegraphics[angle=-90,width=.95\linewidth]{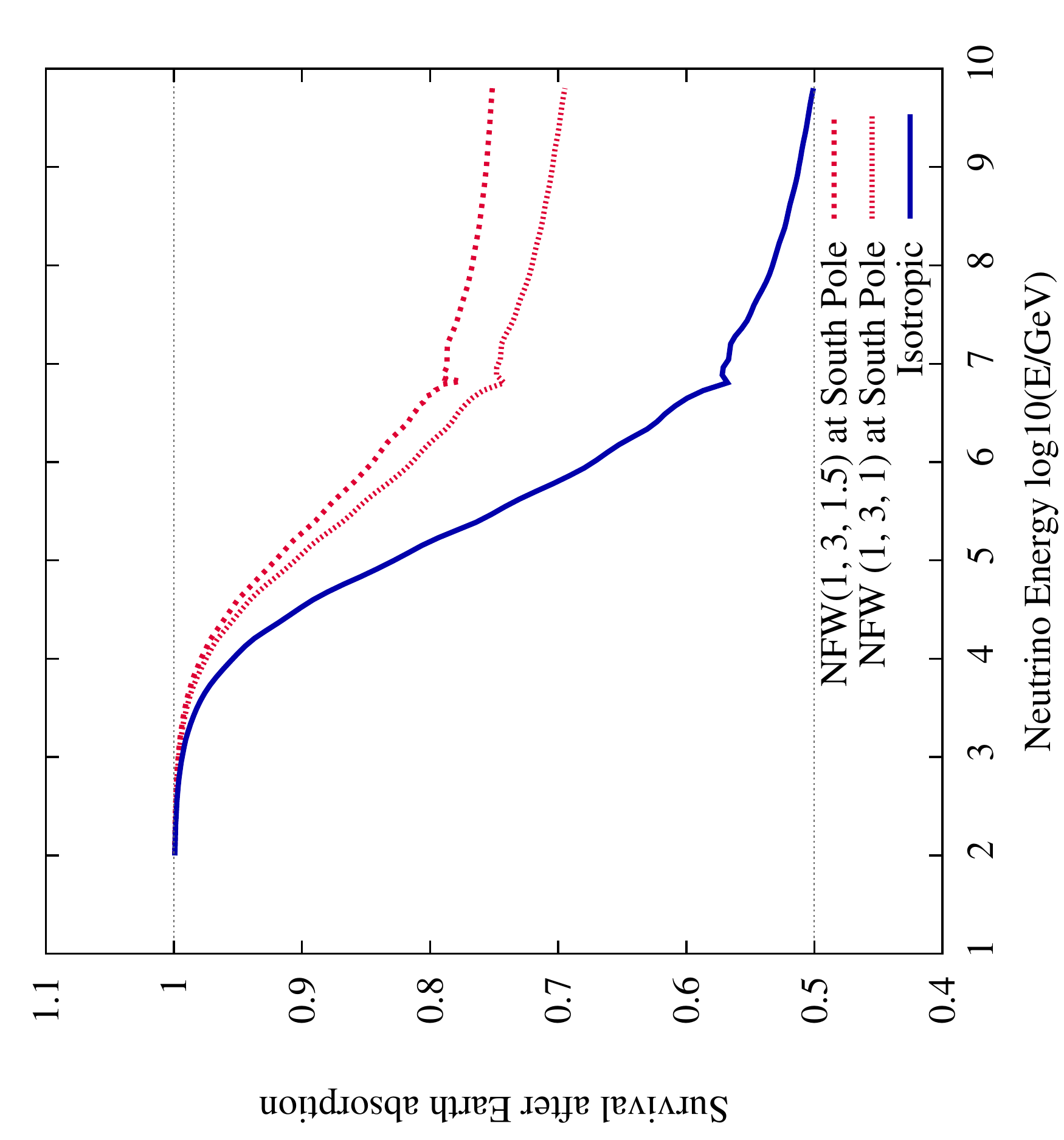}
\end{center}
\caption{Neutrino survival probability due to Earth absorption at the South Pole near surface location as function of the neutrino energy for an isotropic flux compared to the flux expected from dark matter decay originating from a NFW halo distribution and parameterized as~\cite{Yuksel:2007ac}. The effect of the Glashow resonance at 6.3~PeV for anti-electron neutrinos becomes small as we have averaged overall neutrinos flavours. Detailed treatments can be found elsewhere~\cite{Kistler:2013my}. 
 }
\label{fig1}
\end{figure}

For the Galactic signal from $\chi \rightarrow \nu h$ we expect two components, a continuum (from the cascade decay of $h$) and a nearly mono-energetic line at $E_\nu \approx M_\chi/2$. 
To constrain the dark matter lifetime we use the most observable feature of the DM particle decaying with some branching fraction to a neutrino and other particles, which is the neutrino line signal. We assume equal branching ratios to all neutrino flavours and approximate the flux at Earth with 1:1:1 with equal ratio between neutrinos and anti-neutrinos. Using only the Galactic origin we obtain the most model independent bound for any scenario in which neutrinos are produced directly through DM decay. The spectral component from cosmological decays is redshifted to lower neutrino energies and hence can be neglected for our analysis~\cite{Yuksel:2007dr}, however the extra-galactic contribution could be valuable to verify a signal. The expected neutrino flux will further be affected by neutrino absorption in the Earth at high-energies, to include this effect we calculate a survival rate that is shown in Fig.~\ref{fig1}. At a PeV, this effect removes about 20\% of the signal. Also shown is the survival rate for an isotropic flux, which has been assumed when creating IceCube's effective areas used here~\cite{Aartsen:2013jdh}. For our calculation we remove the absorption effect already included in the effective area and apply the survival rate for the neutrino flux expected from the dark matter distribution. Note, that a full correction is not possible as the effective areas released by IceCube are averaged for neutrino and anti-neutrino fluxes. However the impact on our final result is small and can hence be neglected.

The most detectable neutrino signal will be at the highest energy, at this point atmospheric backgrounds are smallest and the signal will be most present. Since IceCube has reported results as function of the deposited energy, which corresponds approximately to the neutrino energy for an electromagnetic shower of an electron neutrino, we here concentrate only on cascade events from $\nu_e$ and $\nu_\tau$
as our signal. By including other flavours and neutral current interactions, which will produce signals at lower electromagnetic equivalent energies, one can further improve in sensitivity, but for our analysis that is focused on the neutrino from Higgs decay, they are not relevant. Muon event contributions to the background estimate given by IceCube that we use to derive our limit only become relevant for energies below 100~TeV and hence we constrain ourselves to the region where the contribution from muon events are negligible.

 We compute the expect event rates from charged current neutrino interactions of electron and tau flavour and compare them to the reported observed events and expected atmospheric backgrounds of the equivalent electromagnetic energy. Neutral current interactions have a much smaller cross section and events are not considered as their electromagnetic equivalent energy would be smaller than the neutrino energy~\cite{Laha:2013lka}. Muon neutrino events have deposited energies significantly lower than the neutrino energy and are hence also not considered. To be conservative we do not subtract any astrophysical neutrino flux, inclusion of such a flux would only make our limits stronger and hence our analysis is more conservative as we underestimate the background. For tau neutrino events the equivalent electromagnetic energy strongly depends on the decay mode of the tau. \red{It is broadly distributed and on average equals approximately half that of the tau neutrino energy~\cite{Kowalski}. A full simulation would be required to determine the exact distribution of the observed electromagnetic equivalent energy at the IceCube detector~\cite{Aartsen:2013vja} and is beyond the scope of this work.  We hence adopt the approximation that the average the deposited energy is roughly half of the neutrino energy~\cite{Kistler:2013my,Dutta:2015dka} and hence approximately half of the tau neutrino events would fall into the two energy bins defined by the tau neutrino energy and the next lower energy bin. As a reminder the IceCube analysis energy bins spread 0.2 in log of the energy. Hence, half of the neutrino energy of $E$ at the centre of an energy bin would correspond to the lower edge of the adjacent lower energy. We point out that a full analysis should be performed by the IceCube collaboration taking the reconstructed energies and corresponding uncertainties on an event by event basis into account. We here only attempt to get an approximation of the bound. Following our assumption, tau neutrino events will largely be contained in the two adjacent energy bins of the IceCube analysis, we assume that 50\% of the tau neutrinos are observed in these bins. We then compare the expected signal flux to the sum of observed events of the corresponding energy bin and the next lower bin. }

The expected number of neutrino events per flavour is given by
\begin{equation}
N = \frac{1}{\tau}  J_{4\pi} \frac{R_{\rm sc}\rho_{\rm sc}}{4\pi m_{\chi}}  4\pi A_{\rm eff}(E=m_{\chi}/2) T_{\rm life} \frac{N_{\nu}}{3},
\label{dphidE_decay}
\end{equation}

where $R_{\rm sc}$ and $\rho_{\rm sc}$ are scale factors~\cite{Yuksel:2007ac}, $m_{\chi}$ dark matter particle mass, $A_{\rm eff}$ the neutrino affective area of the corresponding flavour, $T_{\rm life}$ the lifetime of the experiment. $J_{4\pi}$ is the angle average line-of-sight integral over the dark matter density distribution per solid angle. $N_{\nu}$ is the average number of neutrinos produced at the line signal per DM decay. For the assumed branching fraction of 100\% into $\chi \rightarrow \nu h$, $N_{\nu}$ is one. The factor 1/3 indicates the fraction of each neutrino flavour. We use the neutrino flux from the Milky Way halo assuming an NFW profile ($J_{4\pi} \approx 2.0$)~\cite{PalomaresRuiz:2007ry}.

\begin{figure}[t]
\includegraphics[angle=-90,width=.95\linewidth]{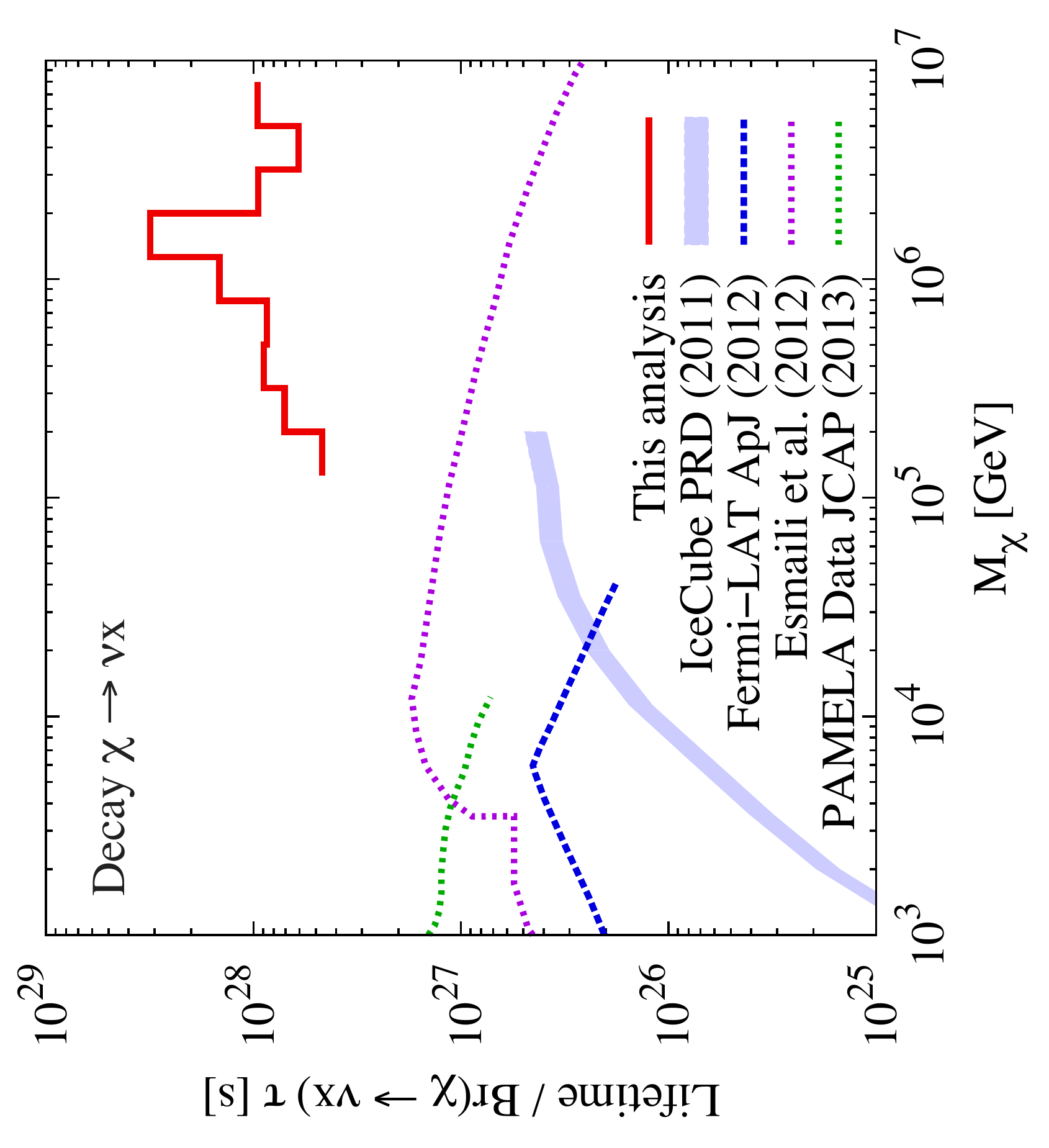}
\caption{Derived limit using the high-energy neutrino flux observed by IceCube in comparison to the previous experimental constraints from IceCube, Fermi-LAT and PAMELA and derived limits from neutrino data~\cite{Esmaili:2012us}. Excluded are regions below the pictured lines. The decay $\chi \to \nu x$ includes $\nu Z$ and $\nu H$ channels thanks to  the Goldstone equivalence theorem. 
\label{fig:derived_limit}}
\vspace{-1.\baselineskip}
\end{figure}

We compute a 90\% C.L. limit on the number of signal events, $N_{90}$, using the observed events and expected background. The observed events and background is \red{taken} as the sum of the bins of $M_{\chi}/2$ and the adjacent lower bin and compare it to the expected neutrino event numbers for a specific decay time. As background estimate we use the prediction from IceCube, including cascade \red{and} track events. The limit is then obtained by $\tau_{90}=\tau \cdot \frac{N}{N_{90}}$. Figure~\ref{fig:derived_limit} shows our derived \red{bound}, following IceCube event binning in neutrino energy~\cite{Aartsen:2014gkd} in comparison to previous limits from the partially instrumented IceCube detector~\cite{Abbasi:2011eq} which investigated the decay of DM into two neutrinos. Note, that the large improvement of our derived limit to the IceCube collaboration result is dominated by the fact that we make use of the neutrino energy, justified by the good energy resolution for cascade events, which is typically better than 15\%~\cite{Aartsen:2014gkd}. The IceCube collaboration analysis relied on the partially instrumented detector and used the up-going muon neutrino event sample and performed a counting experiment of total number of tracks in signal region closer to the Galactic centre compared to a background region. The increase in sensitivity can be simply understood by the fact that the IceCube analysis was not sensitive to neutrino energies as it just counted muon neutrino induced tracks. This counting experiment observed 1389~events in the off-source region and 1367~events in the on-source region, consistent with the null hypothesis. In our analysis we are sensitive to neutrino energies by exploiting contained cascades events. As such we can hence compute the $N_{90}$ energy binwise. The $N_{90}$ in this analysis is closer to two, compared to about 50 in the IceCube halo analysis, hence a factor of twenty improvement at 100 TeV.

Further shown in Fig.\ref{fig:derived_limit} are bounds derived from the Fermi-LAT analysis of gamma-ray emission from the Milky Way halo~\cite{Ackermann:2012rg} and from PAMELA observations of the anti-proton flux~\cite{Cirelli:2013hv} based on the assumed DM decay into $b\bar{b}$. 
The derived limit for the Fermi-LAT gamma-ray line search is justified as $b\bar{b}$ is the dominant Higgs decay channel and further the gamma-ray yield from WW is similar.  Overall our neutrino bound is conservative with respect to the gamma-ray limit as $b\bar{b}$ would result in the strongest limit from gamma-rays.  The observed three PeV neutrinos are seen as `dip' in the two bins covering masses 2-5~PeV in the limit plot as the flux shows `excess' over the expectation. The excess needs further investigation but an extremely interesting interpretation would be the signal from DM. We would invite more dedicated study for further clarification. A complete analysis could further benefit from the less dominant extra-galactic redshifted line spectrum smeared to lower neutrino energies and a potential continuum neutrino spectrum from secondary particle decays. A dedicated IceCube collaboration analysis will be able to improve significantly on our derived limit or lead to the identification of a signal with higher statistics.

\section{Conclusion}
\label{sec:conclusion}

Heavy decaying dark matter might be most detectable with high-energy neutrinos. We use the recently reported observation of high energy extraterrestrial neutrinos, which includes three PeV-energy neutrinos, to set the most stringent bound on 100~TeV to PeV regime. This limit can be achieved by exploiting the line feature present in models with a non zero branching fraction into $\chi \rightarrow \nu + X$. We use the IceCube released data, effective areas and compute neutrino absorption effects in the Earth to derive the constraint. 

Our bound is very conservative in its derivation and suggests that dedicated searches by the experiments can surpass the limit of $10^{28}$~s and have significant discovery potential.

\begin{acknowledgements} 
We would like to thank John Beacom, Matthew Kistler and Shigeki Matsumoto for discussions.
KK is supported in part by Grant-in- Aid for Scientific research
from the Ministry of Education, Science, Sports, and Culture (MEXT),
Japan, Nos. 23540327, 26105520 and 26247042, and by the Center for the
Promotion of Integrated Science (CPIS) of Sokendai (1HB5804100).
SC and CR are supported by Basic Science Research Program through the National Research Foundation of Korea funded by the Ministry of Science, ICT $\&$ Future planning NRF-2013R1A1A2064120 (SC) and NRF-2013R1A1A1007068 (CR).
\end{acknowledgements}


\end{document}